\documentclass{article}

\pdfoutput=1
\usepackage{fullpage}
\usepackage[utf8]{inputenc}
\usepackage{graphicx}
\usepackage{amssymb}
\usepackage{amsmath}
\usepackage{amsfonts}
\usepackage{color}
\usepackage[pdftex]{hyperref}
\usepackage[english]{babel}
\usepackage{subfigure}
\usepackage{algorithm}
\usepackage{algorithmic}
\usepackage{tabularx,colortbl}
\usepackage[sort&compress,numbers]{natbib}
\usepackage{authblk}

\definecolor{Orange}{rgb}{1,0.64,0}
\definecolor{lgray}{rgb}{0.9,0.9,0.9}

\makeindex

\begin{document}

\title{A generative model for protein contact networks}

\author[1]{Lorenzo Livi\thanks{llivi@scs.ryerson.ca}\thanks{Corresponding author}}
\author[2]{Enrico Maiorino\thanks{enrico.maiorino@uniroma1.it}}
\author[3]{Alessandro Giuliani\thanks{alessandro.giuliani@iss.it}}
\author[2]{Antonello Rizzi\thanks{antonello.rizzi@uniroma1.it}}
\author[1]{Alireza Sadeghian\thanks{asadeghi@ryerson.ca}}
\affil[1]{Dept. of Computer Science, Ryerson University, 350 Victoria Street, Toronto, ON M5B 2K3, Canada}
\affil[2]{Dept. of Information Engineering, Electronics, and Telecommunications, SAPIENZA University of Rome, Via Eudossiana 18, 00184 Rome, Italy}
\affil[3]{Dept. of Environment and Health, Istituto Superiore di Sanit\`{a}, Viale Regina Elena 299, 00161 Rome, Italy}
\renewcommand\Authands{, and }
\providecommand{\keywords}[1]{\textbf{\textit{Keywords---}} #1}

\maketitle

\begin{abstract}
In this paper we present a generative model for protein contact networks. The soundness of the proposed model is investigated by focusing primarily on mesoscopic properties elaborated from the spectra of the graph Laplacian. To complement the analysis, we study also classical topological descriptors, such as statistics of the shortest paths and the important feature of modularity.
Our experiments show that the proposed model results in a considerable improvement with respect to two suitably chosen generative mechanisms, mimicking with better approximation real protein contact networks in terms of diffusion properties elaborated from the Laplacian spectra.
However, as well as the other considered models, it does not reproduce with sufficient accuracy the shortest paths structure.
To compensate this drawback, we designed a second step involving a targeted edge reconfiguration process.
The ensemble of reconfigured networks denotes improvements that are statistically significant.
As a byproduct of our study, we demonstrate that modularity, a well-known property of proteins, does not entirely explain the actual network architecture characterizing protein contact networks.
In fact, we conclude that modularity, intended as a quantification of an underlying community structure, should be considered as an emergent property of the structural organization of proteins.
Interestingly, such a property is suitably optimized in protein contact networks together with the feature of path efficiency.\\
\end{abstract}

\section{Introduction}
\label{sec:intro}

Protein contact networks (PCNs) are minimalistic models of protein 3D structures, which collapse the full-rank information of 3D coordinates of each atom into an adjacency matrix providing the pairwise contacts between residues \cite{yan2014construction,doi:10.1021/cr3002356,csermely2012disordered,tasdighian2013modules,randic2011graphical,C4CP01131G,ecoli_graph__arxiv,ecoli_graph_complexity_arxiv,mixbionets1__arxiv,mixbionets2}.
A contact is scored if the Euclidean distance between alpha carbons of each residue pairs is within two van der Waals radii.
PCNs allow for a reliable reconstruction of the global protein structure \cite{vassura2008reconstruction}.
Moreover PCNs allow for an efficient description of relevant biological properties of proteins such as allosteric effect and identification of active sites \cite{pcn_typology2015}.
The efficiency of PCNs in retaining the essential features of protein structures makes the development of a PCN generative model of utmost importance for shedding light on both folding process and the structural bases of the unique functional properties of protein molecules \cite{Finkelstein200423,Wolynes2014,banavar2007physics,orozco2014theoretical}.
Although many generative models have been developed in the still young network science discipline \cite{newman2003structure,boccaletti+latora+moreno+chavez+hwang2006}, fewer and less established examples are available in the literature when focusing on formal representations of protein molecules \cite{estrada2010universality,bartoli2007effect,balakrishnan2011learning,mishra2012knot,penner2010fatgraph,penner2011algebro,boomsma2008generative,segal2009novel}.
It is possible to cite approaches for modeling proteins based on knot theory \cite{mishra2012knot} and topology \cite{segal2009novel,penner2010fatgraph,penner2011algebro}.
Other approaches focus on approximating particular protein structures, such as the local structure \cite{boomsma2008generative} and specific fold families \cite{balakrishnan2011learning}.
To the best of our knowledge, only Refs. \cite{estrada2010universality,bartoli2007effect} studied the problem of generating PCNs for evaluating detailed graph-theoretical characteristics.

The quest for a reliable and, most importantly, justifiable generative model for PCNs implies as a first step the identification of a target function. This allows for a unambiguous evaluation of the proposed model in terms of ``superposition'' of the simulated contact networks with the real PCNs.
Inspired by the seminal works by \citet{doi:10.1146/annurev.physchem.59.032607.093606}, we considered here as target function to approximate of the peculiar heat trace decay of PCNs \cite{mixbionets1__arxiv}.
Such a property is elaborated from the heat kernel \cite{Xiao:2009:GCH:1563046.1563099,chung2007heat,kloster2014heat}, the graph-theoretical analogue of the well-known first-order differential equation describing diffusion of heat in a physical medium.
The heat trace (HT) is an invariant property -- in the graph-theoretical sense -- elaborated from the spectrum of normalized graph Laplacian \cite{PhysRevE.77.031102,mitrovic2009spectral,Merris_1994,almendral2007dynamical}.
Graph Laplacians are objective of focused studies in many scientific fields, as in fact it is possible to extrapolate many structural and dynamical properties from such a matrix representation of the system \cite{Peixoto_2013,zhang2014spectra,kuhn2011spectra,de2013laplacian}.
The interest in the study of graph Laplacians motivated also related researches focusing on the so-called spectral reconstruction, i.e., on calculating a graph given a specific spectrum to be considered as target \cite{Ipsen2002,Comellas_Diaz-Lopez_2008}. Although potentially interesting, co-spectrality of graphs is a very hard problem and it is still a not very well-developed field in graph theory \cite{van2009developments}, limiting hence its practical exploitation.

Inside a protein molecule, energy readily flows between distant regions connected by inter-module contacts (fast lane) and only slowly within modules (slow lane).
In other words, proteins present either a strong modularity, causing a low thermal dissipation (heat is kept into modules by the richness of dead ends pathways slowing down the spreading of energy), and a suitable number of long-range shortcuts, allowing for a rapid and yet efficient communication between distant sites responsible for allosteric effect \cite{doi:10.1146/annurev.physchem.59.032607.093606}.
This dual behavior is at the basis of two crucial properties for protein physiology: 1) keeping a stable micro-environment for the catalyzed reaction (slow decay) and 2) allowing for an efficient spreading of information throughout the molecule to get rid of environmental changes (binding of an allosteric effector, pH changes, etc).
There is a clear trade-off between the two above goals: increasing modularity is good for the first goal but is detrimental for the latter; on the other hand characteristic length minimization is an efficient strategy for the latter but is detrimental for the former.

The contribution of this paper consists in a two-step generative model for PCNs; the first stage of our method takes inspiration from the work of \citet{bartoli2007effect}.
The dataset considered in our study consists of four ensembles (classes) of networks: i) actual PCNs elaborated from the E. coli proteome \cite{ecoli_graph__arxiv,ecoli_graph_complexity_arxiv}, ii) synthetic networks generated according to the recipe of \citet{bartoli2007effect}, iii) synthetic modular networks generated with the method proposed by \citet{communitystructure_bmc2014}, and finally iv) those generated with our method.
We evaluate the soundness of the proposed approach by focusing on mesoscopic analyses. Notably we first study characteristics elaborated from the normalized Laplacian spectra.
To complement this spectral analysis, we also study topological descriptors, including statistics of the shortest paths and quantification of the network modularity.
Results show that the ensemble of networks generated with our method ends up in a statistically significant improvement of similarity with real PCNs, as for both spectral and topological properties.
However, a principal component analysis (PCA) of the considered topological descriptors revealed a gap with actual PCNs, specifically related to the shortest paths. 
The second step of the proposed method is hence designed to compensate this drawback.
A new version of the ensemble of networks is then obtained by rewiring edges with high edge-betweenness \cite{cuzzocrea2012edge}. 
As a result, we show that we are able to achieve a further statistically significant improvement of the ensemble characteristics, without altering the global spectral properties of the first ensemble that we generated.

As a byproduct of our study, we demonstrate that modularity, a well-known feature found in proteins as well as in many other biological networks, is not sufficient to explain the underlying network architecture of PCNs. This result is of particular interest, since it stresses the peculiar architecture of proteins that suitably merges conflicting features such as path efficiency and modularity.
The topological properties of the obtained graphs, exploited by PCA, allowed us to offer a clear structural counterpart to HT features.
These structural counterparts are by no means confined to protein science, given that well-known relations between small-world, fractal, and modular architectures with path efficiency \cite{gallos2012small,rozenfeld2010small} are present also in brain networks \cite{bullmore2012economy} and are likely to constitute relevant features to be optimized in other natural and/or artificial networked systems.

The remainder of this paper is structured as follows. In Section \ref{sec:dataset} we describe the data that we considered in our study.
In Section \ref{sec:results} we show and discuss the obtained results.
Section \ref{sec:conclusions} offers the conclusions and pointers to future research works.
The proposed generative algorithm is fully described in the Methods section, notably in Section \ref{sec:our_method}; the considered graph characterizations are instead introduced in Section \ref{sec:graph_characterization}.

\section{Dataset}
\label{sec:dataset}

In our study we consider four ensembles of networks, three of which are created according to specific generative models.
Each ensemble contains 100 networks of varying size (from 300 to 1000 vertices). Each graph with $n$ vertices and $m$ edges is present in the four different ensembles preserving $n$ and $m$, but varying the resulting topology according to a generative mechanism.
This is performed to focus the analysis on ensemble features proper of the structural organization, without considering effects due to the size of the graphs.

\begin{figure}[ht!]
 \centering
 \includegraphics[viewport=0 0 342 243,scale=0.6,keepaspectratio=true]{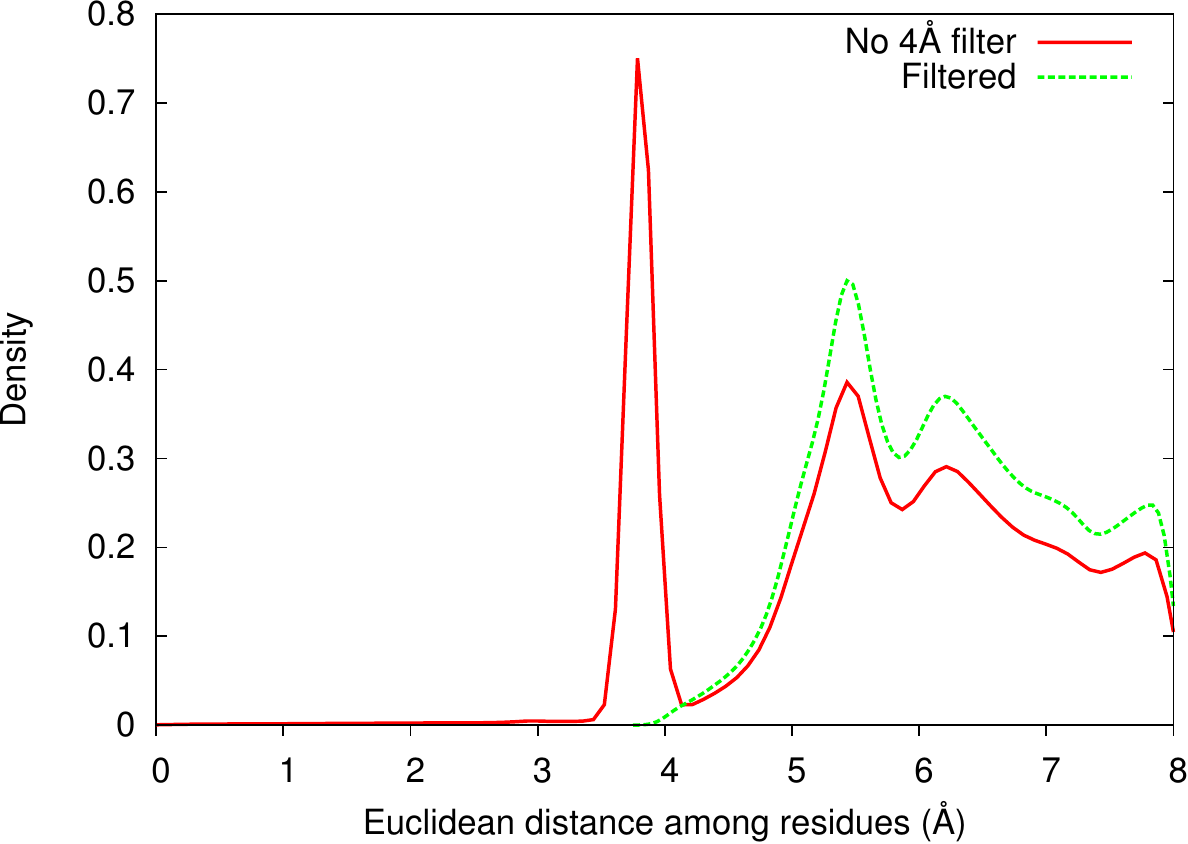}
 \caption{Density of Euclidean distances among native contacts in PCNs.}
 \label{fig:cld}
\end{figure}

The first ensemble of graphs contains PCNs, directly obtained from the 3D native structures resolved for the E. coli proteome \cite{ecoli_graph__arxiv,ecoli_graph_complexity_arxiv}. Each vertex is defined as the alpha carbon of a residue; edges are added among two residues if their Euclidean distance is within the $[4, 8]$ \AA\ range. This choice is justified by observing the densities of native contact length in Fig. \ref{fig:cld}, respectively when considering all contacts below 8 \AA\ and the herein adopted filtered version.
It is possible to note a peak right before 4 \AA, which corresponds to trivial contacts due to closeness of the residues on the backbone \cite{doi:10.1021/cr3002356}; such contacts do not alter the typical network architecture of PCNs and therefore they are not considered in our PCN graph representation.

The second ensemble of networks is elaborated by using the recipe of \citet{bartoli2007effect}, which considers a set of edges added deterministically due to the backbone and additional edges inserted according to a probability that scales linearly with the sequence distance of residues.
For the third ensemble we use networks generated according to the recently-proposed scheme by \citet{communitystructure_bmc2014}. Such a generation mechanism produces modular networks with controlled (i.e., used-defined) modularity and degree values.
Such a generation mechanism is inspired by the fact that modular structures seem to be ubiquitous in biological networks; modularity is considered to be at the basis of resilience and adaptability features of biological networks.
It is well-known that PCNs are modular, i.e., they posses a well-defined community structure \cite{mixbionets1__arxiv}. Accordingly, the third ensemble of Sah et al. networks is generated by copying modularity and degree from the considered PCNs.
The contemporary presence of networks in which modularity (if any) comes up as an emergent property and Sah et al. networks, in which modularity is a built-in property, will help us to check if the peculiar PCN spectral and topological properties are a direct consequence of their pronounced modularity or not.
Finally, the fourth ensemble of networks is constituted by 100 graphs generated according to the herein proposed mechanism, referred to as LMGRS networks, whose first step consists in a variant of the Bartoli et al. model. In this adaptation, the linear scaling of the probability of the edges with the distance in sequence is replaced by the empirical frequencies measured in the PCN ensemble.
As it is discussed later, the LMGRS are successively reconfigured in a iterative fashion to obtain a new ensemble of networks, shortened in the following as LMGRS-REC. The proposed LMGRS generation and LMGRS-REC reconfiguration methods are fully described in Sec. \ref{sec:our_method}.

\section{Results}
\label{sec:results}

\begin{figure}[ht!]
\centering

\subfigure[HT slopes decay.]{
\includegraphics[viewport=0 0 351 241,scale=0.62,keepaspectratio=true]{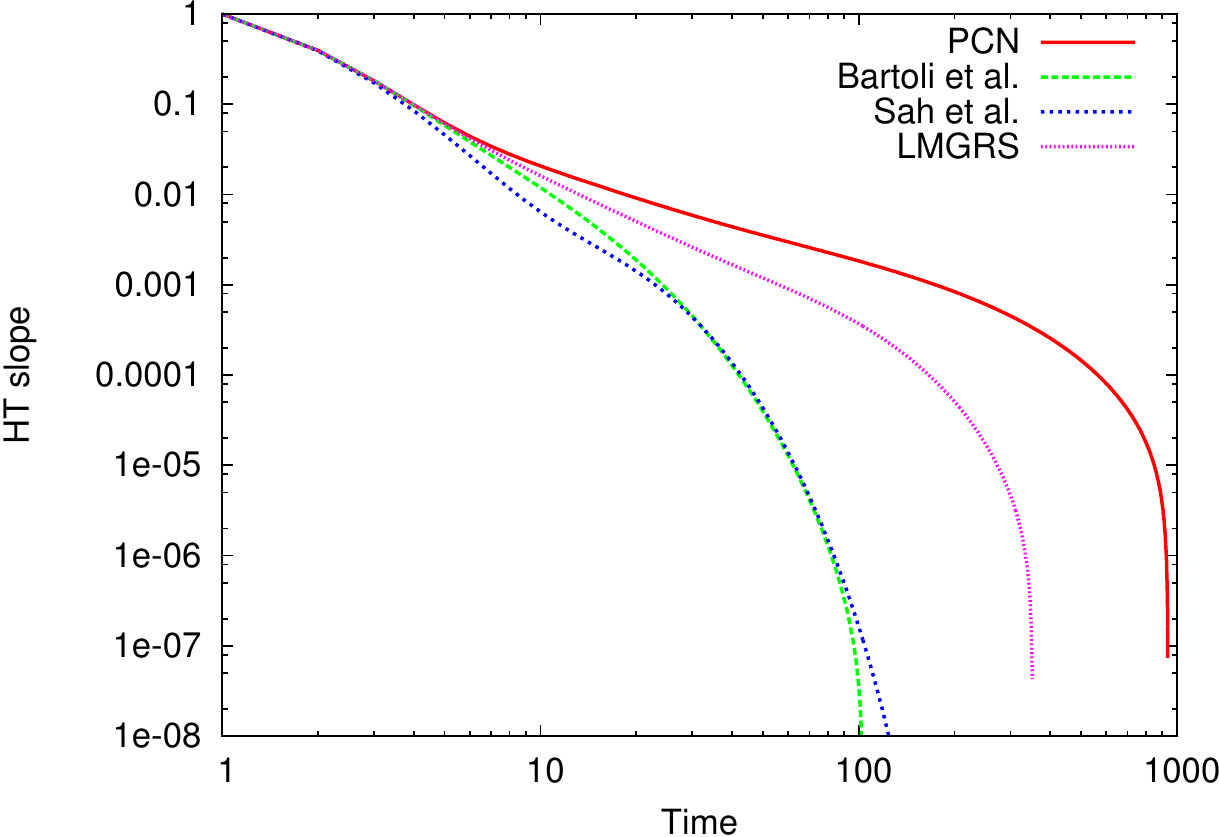}
\label{fig:HT_slopes}}
~
\subfigure[Spectral densities of ensembles.]{
\includegraphics[viewport=0 0 342 243,scale=0.62,keepaspectratio=true]{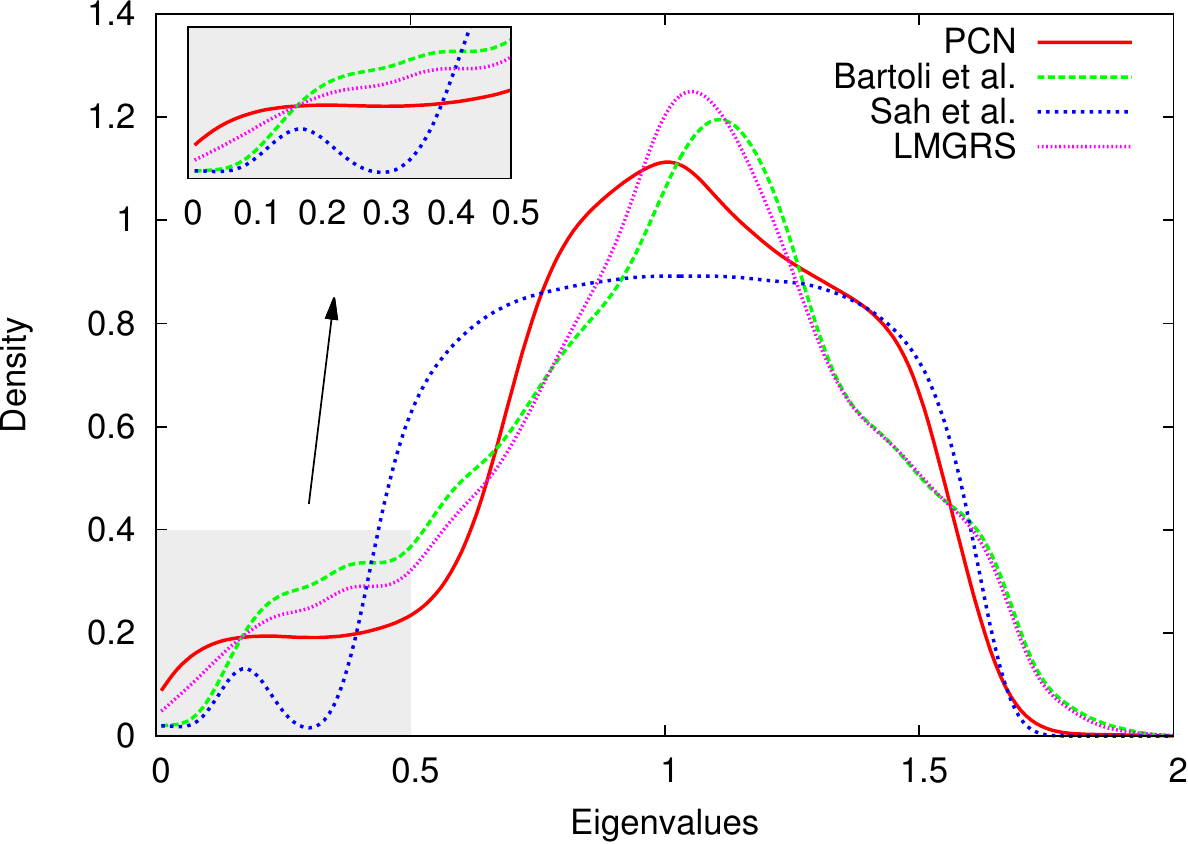}
\label{fig:spectral_densities}}

\caption{Ensemble HT slopes decay (\ref{eq:ensemble_ht}) for the considered graphs \ref{fig:HT_slopes} and related Laplacian spectral densities \ref{fig:spectral_densities}. The proposed LMGRS model clearly denotes more similar characteristics with respect to PCNs in terms of HT decay. Analogously, the LMGRS model induces a density of eigenvalues more similar to PCNs in the lower bands (see detailed plot), suggesting that the community structure is more suitably approximated. The Sah et al. model instead does not mimic the spectral density, even if the modularity level and degrees have been copied from the original PCNs.}
\label{fig:laplacian}
\end{figure}

The four ensembles are first evaluated in terms of mesoscopic properties elaborated from the spectra of the normalized graph Laplacians (see Sec. \ref{sec:graph_characterization} for technical details).
Fig. \ref{fig:laplacian} shows, respectively, the characteristic HT decay and the ensemble spectral densities.
The HT decay analysis (\ref{fig:HT_slopes}) offers insights on an ensemble in terms of characteristic diffusion time.
The analysis takes into account the varying-size character of the considered networks. From the plot it is possible to note that LMGRS introduces a considerable improvement with respect to Bartoli et al. and Sah et al. ensembles: the former decays around $t\simeq350$ while the latter decay much faster at $t\simeq 100$. However, the PCN trend is not accurately approximated yet.
This improvement of several order of magnitudes in the characteristic HT decay time can be explained by focusing on the spectral densities shown in Fig. \ref{fig:spectral_densities}. LMGRS ensemble density possesses clear similarities with the one of Bartoli et al., being the two based on the same algorithmic template. Nonetheless, by highlighting the lower band of the spectra, it is possible to note some important differences for a specific region (in-between 0 and 0.2) containing eigenvalues related to the modular structure of networks.
LMGRS ensemble offers a better approximation for those small eigenvalues, which explains the significant improvement observed for the HT decay.

\begin{figure}[ht!]
\centering
\subfigure[PC1--PC2.]{
\includegraphics[viewport=0 0 343 245,scale=0.62,keepaspectratio=true]{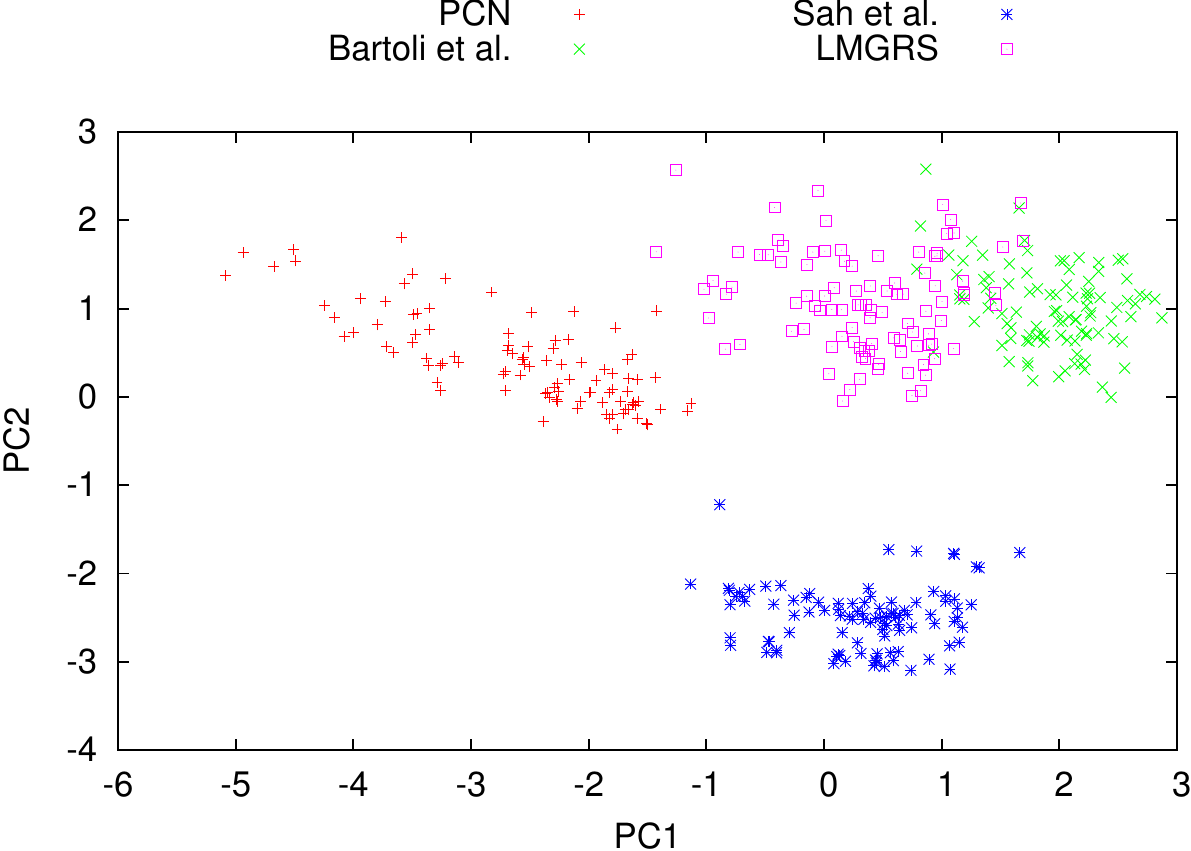}
\label{fig:TD_PCA1-2}}
~
\subfigure[PC1--PC3.]{
\includegraphics[viewport=0 0 343 245,scale=0.62,keepaspectratio=true]{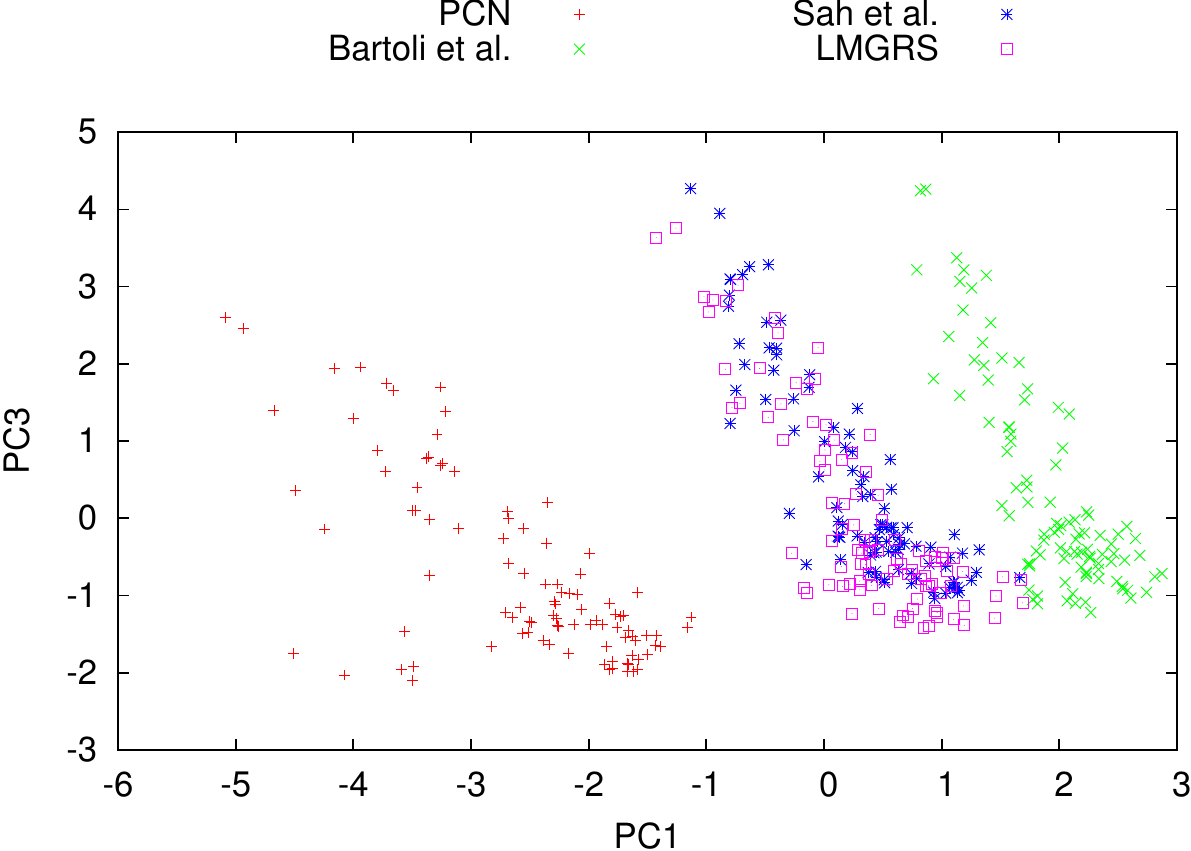}
\label{fig:TD_PCA1-3}}

\subfigure[PC2--PC3.]{
\includegraphics[viewport=0 0 343 245,scale=0.62,keepaspectratio=true]{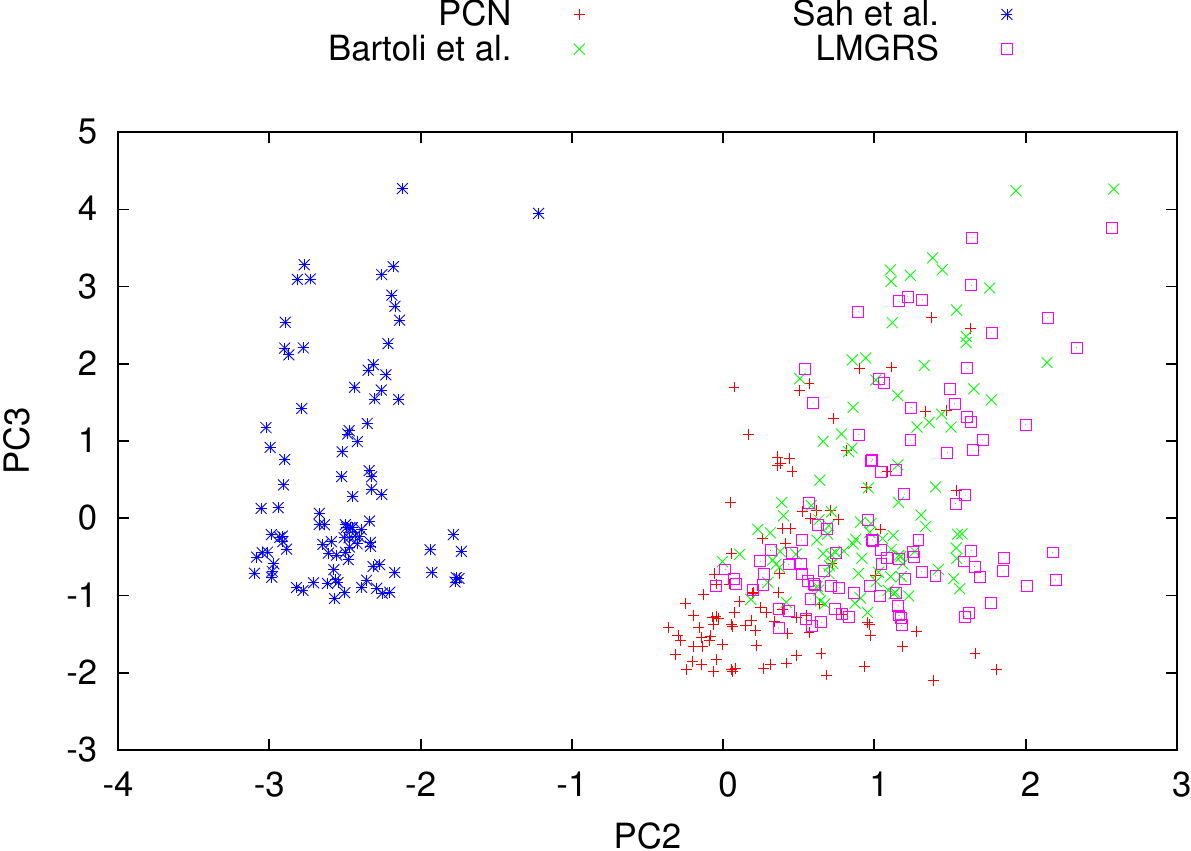}
\label{fig:TD_PCA2-3}}
~
\subfigure[PC1--PC2--PC3.]{
\includegraphics[viewport=0 0 309 182,scale=0.7,keepaspectratio=true]{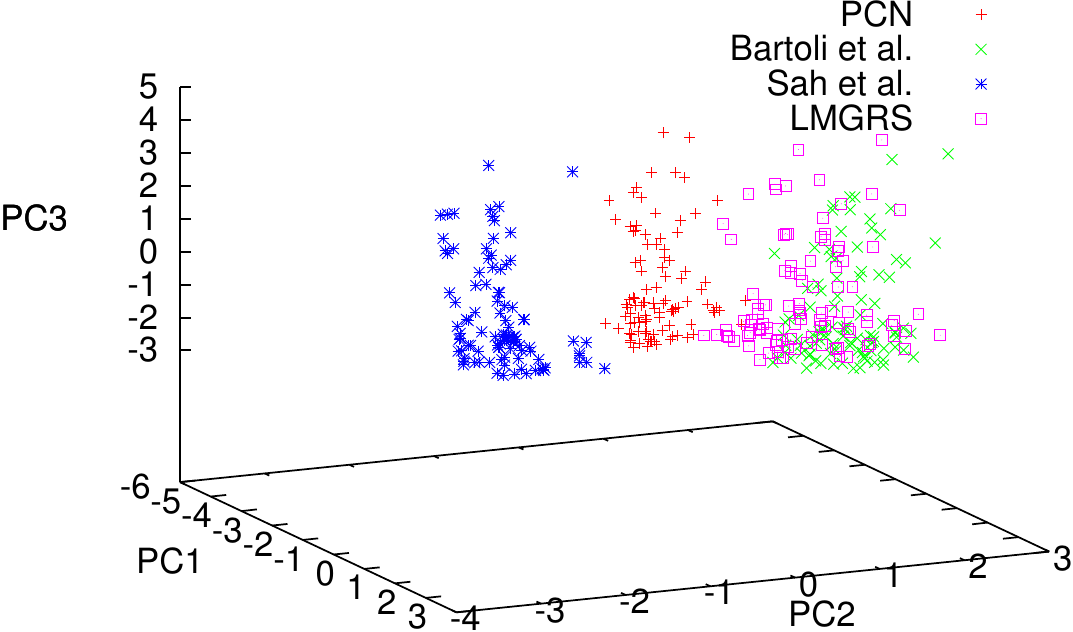}
\label{fig:TD_PCA1-2-3}}

\caption{PCA of the topological descriptors calculated on the four ensembles of protein graphs. LMGRS model is an improvement with respect to \citet{bartoli2007effect} also considering classical TD.}
\label{fig:TD_PCA}
\end{figure}

Now we move to the analysis of the ensembles by considering the representation of each graph as a numeric vector containing suitable features that characterize different aspects of the network topology (see Sec. \ref{sec:graph_characterization}). To offer a synthetic visualization, in Fig. \ref{fig:TD_PCA} we show the pairwise relations among the first three principal components (PC) elaborated via the principal components analysis (PCA) of such a vector-based representation of the considered networks; data is standardized prior to PCA processing.
The first three PCs explain $\simeq 92\%$ of the entire data variance (PC1 $\simeq 39\%$, PC2 $\simeq 30\%$, and PC3 $\simeq 23\%$) and therefore they are considered as the signal part of information; the component loadings (Pearson correlation coefficients between original descriptors and components) are reported in Tab. \ref{tab:factors}.
Loadings on the PC offer a particularly interpretable scenario, where PC1 is mostly explained by the path distribution (ACC and ASP) and the local clustering (ACL). 
As expected ACC negatively scales with both ASP and ACL, so pointing to the fact that ACL decreases the efficiency of signal transmission across the network (positive correlation with ASP). Thus, high values of PC1 corresponds to architectures with elevated characteristic length (slow information transmission), while low values of PC1 point to graphs with elevated closeness centrality (ACC) and thus relatively efficient signal transmission.
PC2 is mainly correlated with MOD, A and H, with A going in the opposite direction with respect to the other two descriptors. This corresponds to the fact the regularity of a graph decreases as the modularity increases; it is also well-known that modularity affects random walks behavior, explaining the positive correlation with H.
PC3 is entirely described by the spectra of the adjacency and Laplacian matrices (respectively indicated by EN and LEN).

The PCs are linearly independent by construction, so the above results clearly indicate that the dataset can be described by three autonomous topological features: 1) path length and local clustering (PC1); mesoscopic modularity (PC2); and 3) spectral properties (PC3).
The particular mixing of these independent features varies across the different ensembles.
Let us now focus on the PCA subspace spanned by PC1--PC2 (Fig. \ref{fig:TD_PCA1-2}). It is possible to note that LMGRS ensemble introduces an improvement in PC1, which as explained before, encodes contributions in terms of path distribution and local clustering. A very interesting scenario can be observed when considering the projection given by PC1--PC3 (Fig. \ref{fig:TD_PCA1-3}). In fact, when PC2 is not considered Sah et al. and LMGRS become very similar to each other, and entirely different from the ensemble of Bartoli et al.
To summarize, it is worth pointing out that the average Euclidean distance among the LMGRS and PCNs networks represented in the three-dimensional PCA space is significantly inferior ($p<0.0001$) with respect to the distances among Bartoli et al. and PCNs (3.13 vs 4.69 with standard deviations 0.75 and 0.68, respectively).
\begin{table}[th!]
\caption{Principal component loadings.}
\begin{center}
\begin{tabular}{|c|c|c|c|}
\hline
\rowcolor{lgray}  & \textbf{PC1} & \textbf{PC2} & \textbf{PC3} \\
\hline
\cellcolor{lgray}\textbf{MOD} & 0.26148 & \textbf{0.78872} & 0.13409 \\ 
\cellcolor{lgray}\textbf{ACC} & \textbf{-0.97835} & -0.15695 & -0.11588 \\ 
\cellcolor{lgray}\textbf{ASP} & \textbf{0.92838} & 0.26494 & 0.09270 \\ 
\cellcolor{lgray}\textbf{ACL} & \textbf{0.89098} & -0.22533 & -0.12227 \\ 
\cellcolor{lgray}\textbf{EN} & 0.01862 & 0.27707 & \textbf{0.95810} \\ 
\cellcolor{lgray}\textbf{LEN} & 0.04354 & -0.27981 & \textbf{0.94213} \\ 
\cellcolor{lgray}\textbf{H} & 0.05171 & \textbf{0.99519} & -0.04393 \\
\cellcolor{lgray}\textbf{A} & 0.09073 & \textbf{-0.84835} & 0.06463 \\
\hline
\end{tabular}
\label{tab:factors}
\end{center}
\end{table}

\begin{figure}[ht!]
\centering

\subfigure[Average absolute differences.]{
\includegraphics[viewport=0 0 340 231,scale=0.62,keepaspectratio=true]{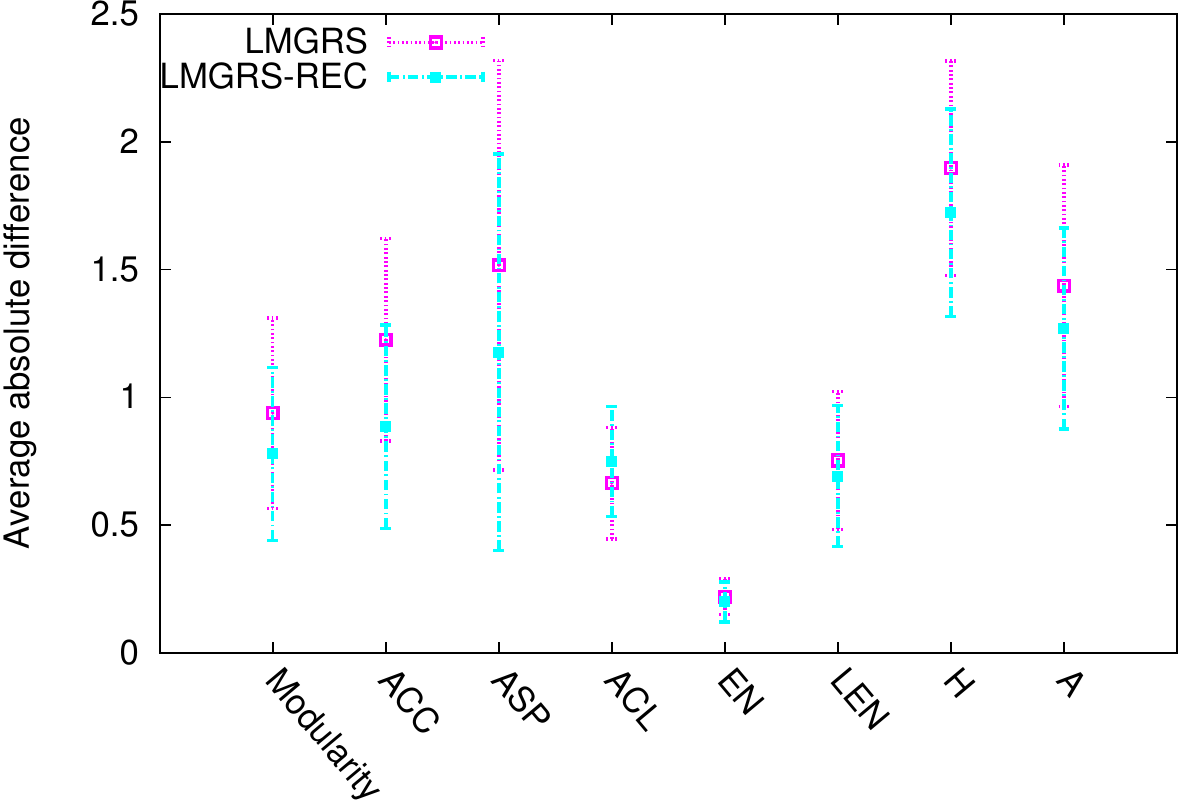}
\label{fig:td_avf_diff}}
~
\subfigure[ASP over iterations.]{
\includegraphics[viewport=0 0 346 241,scale=0.62,keepaspectratio=true]{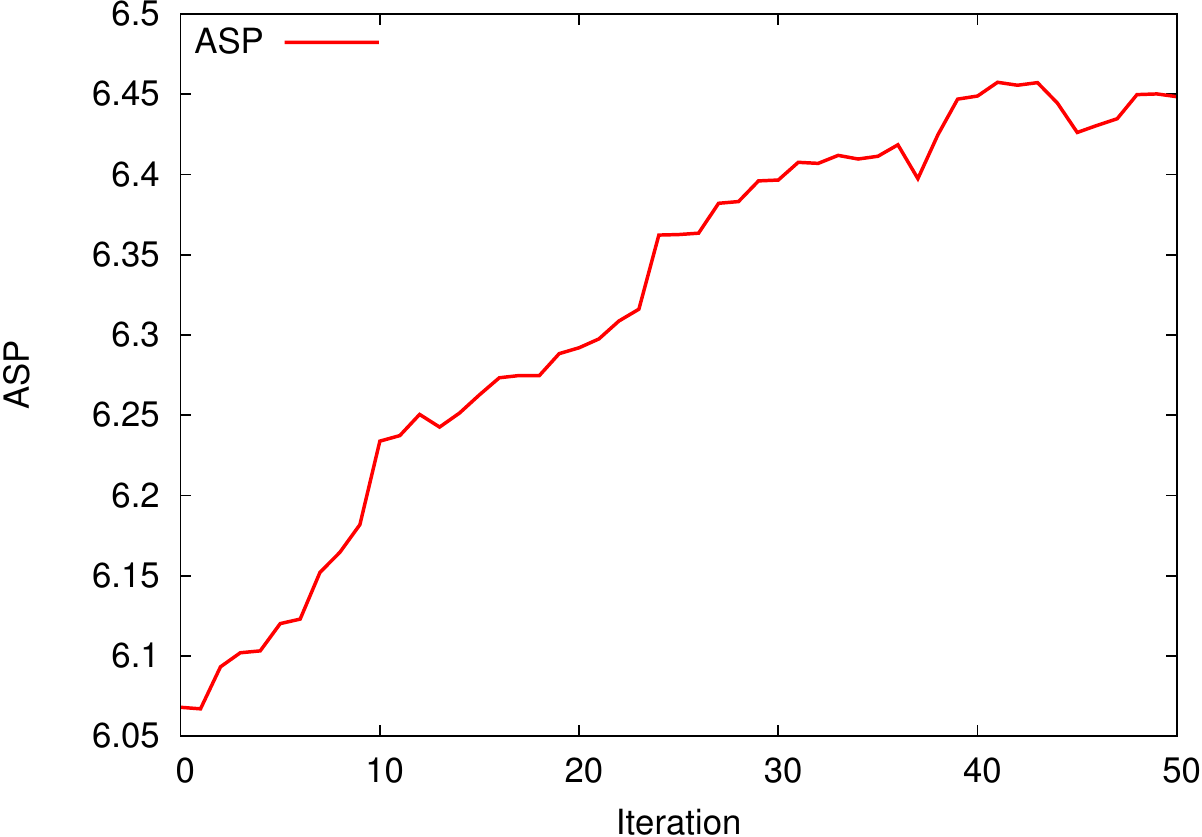}
\label{fig:jw0058_lmgr__ASP}}

\caption{Average differences for each TD with respect to PCN (\ref{fig:td_avf_diff}) and their standard deviations. We are able to modify, among the other factors, the small-world character of the generated networks without significantly affecting the spectra of the adjacency (EN) and Laplacian (LEN) matrices. Statistical significance of differences is assessed via t-test. Fig. \ref{fig:jw0058_lmgr__ASP} shows the ASP of a sample graph during the reconfiguration process.}
\label{fig:td_avf_diff_ASP}
\end{figure}

\begin{figure}[ht!]
\centering

\subfigure[HT slopes.]{
\includegraphics[viewport=0 0 351 241,scale=0.62,keepaspectratio=true]{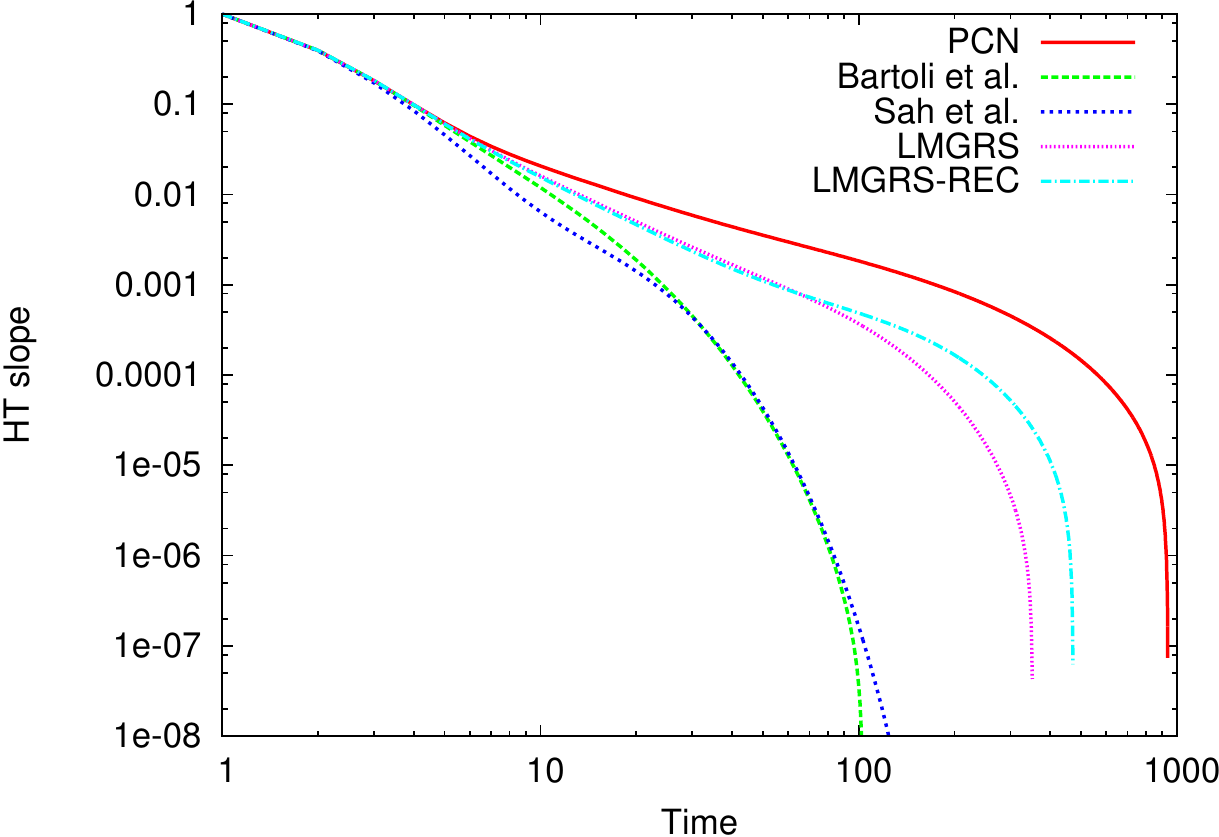}
\label{fig:HT_slopes_rec}}
~
\subfigure[Spectral densities.]{
\includegraphics[viewport=0 0 342 243,scale=0.62,keepaspectratio=true]{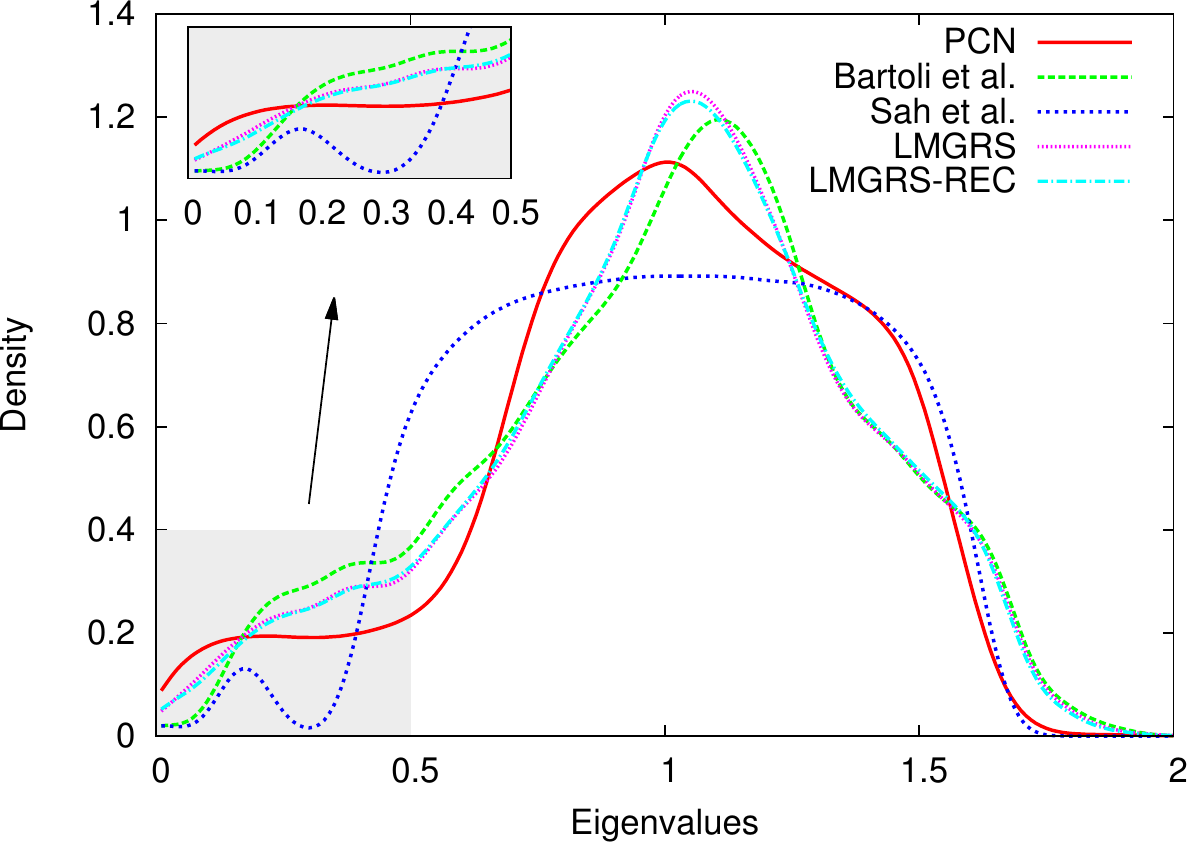}
\label{fig:spectral_densities_rec}}

\caption{Same as Fig \ref{fig:laplacian} but including also reconfigured LMGRS.}
\label{fig:laplacian_rec}
\end{figure}

From PCA space snapshots we deduce that the LMGRS ensemble is a considerable improvement with respect to the others, while there is still a gap to be filled with respect to PCN. Notably, LMGRS networks present a too small characteristic path length -- the small-world signature is too strong.
This fact explains the differences observed for what concerns path distribution and modularity, since in fact the path efficiency and modular properties are two conflicting features in networks, which usually survive in the same network in terms of a trade-off (modular organization is progressively lost as the network becomes more and more efficient in terms of shortest paths).
To this end, as mentioned earlier, we consider another ensemble derived by post-processing LMGRS networks with the edge reconfiguration process described in Sec. \ref{sec:our_method}, denoted as LMGRS-REC.
The reconfiguration process is targeted to rewire edges with high edge-betweenness, since those edges have a direct impact on path efficiency, and accordingly also on the modular organization.
Each graph of the LMGRS ensemble is reconfigured according to the following process (see Sec. \ref{sec:our_method} for details).
At each iteration, the edge with maximum edge-betweenness is removed and it is re-attached to two randomly chosen vertices at a backbone distance given by the empirical distribution shown in Fig. \ref{fig:empirical_distribution}.
This process is repeated until a suitable convergence criteria is met. In our case, we considered a number of iterations (50) that resulted in a statistically significant improvement of the ensemble features that we observe.
In Fig. \ref{fig:td_avf_diff} we show the detailed changes of the TD for the LMGRS and LMGRS-REC with respect to the PCNs.
The figure reports, for each descriptor, the average absolute difference calculated for each graph of the respective ensembles with respect to the PCN graphs; standard deviations are reported as vertical bars.
Results show that the reconfiguration algorithm performs well as for statistical significance of differences with PCN, assessed via t-test with the usual 5\% threshold.
In particular, as desired reconfigured networks denote more similar ASP ($p<0.0025$) and ACC ($p<0.0001$). As expected, such improvements for the shortest paths have a direct influence on the global modularity. In fact, MOD is significantly improved ($p<0.0018$). This is a direct consequence of the fact that path efficiency and modularity are features to be considered in a trade-off. ACL similarity improves as well ($p<0.0059$), denoting a better approximation of the local cluster structure of PCNs.
It is important to note that differences for EN ($p<0.0621$) and LEN ($p<0.1134$) are not statistically significant (especially those for LEN).
This fact tells us that spectral properties of the reconfigured networks are not significantly altered.
Fig. \ref{fig:spectral_densities_rec} offers a visual confirmation of this fact. In fact, the spectral densities for LMGRS and LMGRS-REC reported in the figure are almost identical.
However, it is worth discussing the HT slopes shown in Fig. \ref{fig:HT_slopes_rec}. From the figure, it is possible to notice a slight divergence among LMGRS and LMGRS-REC for large-time instants. This is due to the difference in magnitude of the first non-zero eigenvalue of the normalized Laplacian, which particularly influences the asymptotic HT behavior.
Such a difference is a byproduct of the designed edge reconfiguration algorithm, which focuses on rewiring edges with high edge-betweenness: those are most likely connections among different densely connected communities. In graph-theoretical terms, LMGRS-REC networks are characterized by a lower conductance with respect to the LMGRS ensemble. In fact, the conductance is directly related to first non-zero eigenvalue of the Laplacian (details not shown here).
Finally, differences for both H ($p<0.0030$) and A ($p<0.0069$) are significant as well.

\begin{figure}[ht!]
\centering

\subfigure[PC1--PC2.]{
\includegraphics[viewport=0 0 346 245,scale=0.62,keepaspectratio=true]{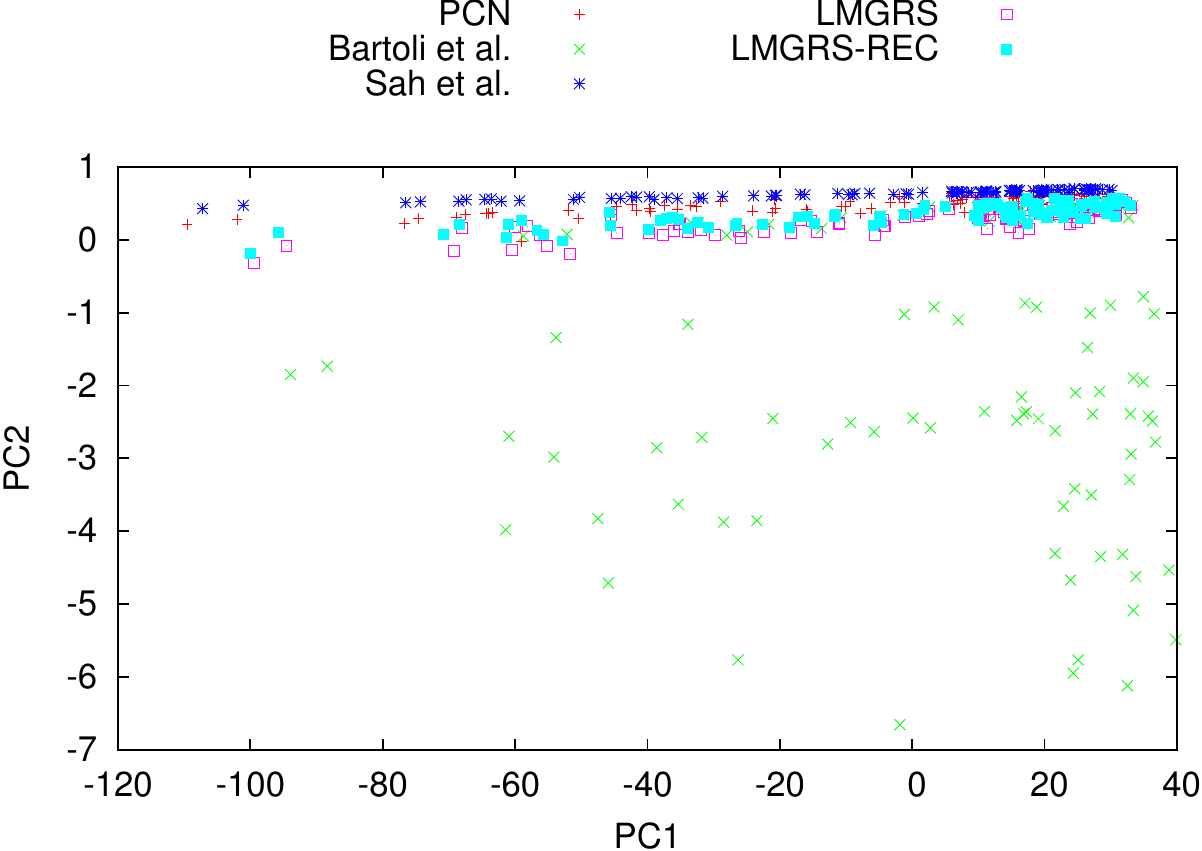}
\label{fig:HC_PCA1-2}}
~
\subfigure[PC2--PC3.]{
\includegraphics[viewport=0 0 346 245,scale=0.62,keepaspectratio=true]{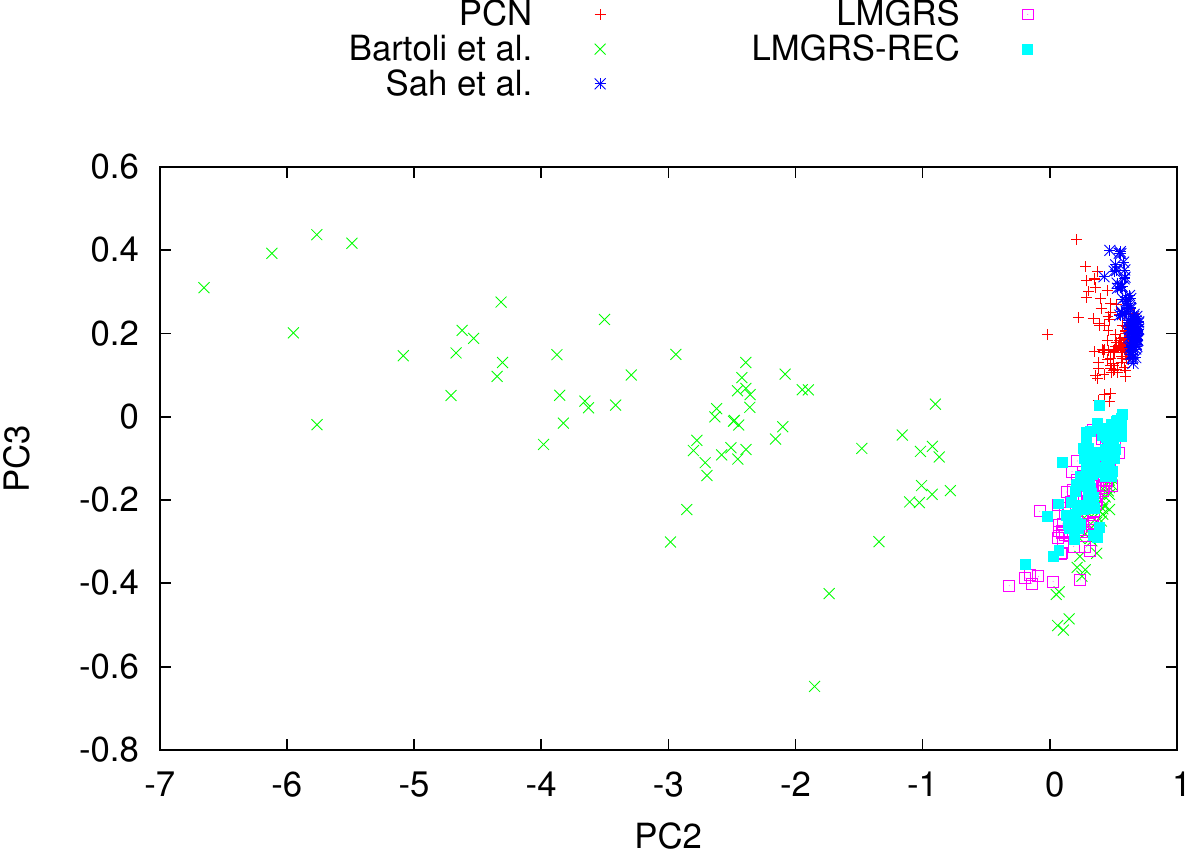}
\label{fig:HC_PCA2-3}}

\caption{PCA of the first 100 time instants of the HC coefficients elaborated from the ensembles.}
\label{fig:HC_PCA}
\end{figure}

To conclude this section, we discuss the results shown in Fig. \ref{fig:HC_PCA}. The figure shows the PCA performed by considering the (standardized) HC coefficients (\ref{eq:heat_content}) of the first 100 time instants.
HC coefficients include, in addition to the spectra, also the information of the eigenvectors of the normalized Laplacian, which encode the arrangement of the vertices in a given vector space. In the literature (see Ref. \cite{mitrovic2009spectral} and references therein) this is called localization of the eigenvectors and it is usually exploited for clustering purposes.
While variance is almost entirely explained by PC1 ($\simeq 99.8\%$), for our purposes we consider the first three PCs.
From PC1--PC2 it is possible to note that PCN, LMGRS, LMGRS-REC, and Sah et al. have a very well-defined and compact configuration in the PCA subspace.
Since Sah et al. posses a well-defined community structure, we deduce that when considering the information of the eigenstructure of the normalized Laplacians, all four ensembles (i.e., PCN, LMGRS, LMGRS-REC, and Sah et al.) posses striking similarities.
However, Bartoli et al. ensemble denotes a very disorganized configuration regardless of the considered PCA subspace (i.e., PC1--PC2 or PC2--PC3), which could be intended as a symptom of a weaker modular organization of the vertices.

\section{Conclusions and future directions}
\label{sec:conclusions}

In this paper, we have proposed a two-step generative model for protein contact networks.
For the first step, we partially took inspiration from the work of \citet{bartoli2007effect}, whose idea was to generate contact graphs by first adding backbone contacts deterministically (considering adjacent residues along the sequence). Successively, a number of additional contacts were added with a probability inversely proportional to the residue distance along the sequence.
Here we modified this part by considering the actual empirical probability distribution of contacts with respect to the sequence distance, derived from an ensemble of E. coli proteins.
We analyzed our generative method by considering three additional ensembles composed of 100 varying-size protein contact networks.
We focused on a mesoscopic analysis, that is, we primarily investigated the soundness of the models by considering information derived from the eigendecomposition of the normalized Laplacian. Results showed that the proposed method approximates with better precision the behavior of actual protein contact networks in terms of characteristic diffusion time.
We considered also several common topological descriptors. This last analysis pointed out that our method, as well as the others, does not approximate sufficiently well the path distribution.
To this end, we designed an edge reconfiguration algorithm to be used as the second step of the proposed generative method.
We then generated an additional ensemble of reconfigured networks, which showed statistically significant improvements with respect to the initial one.

We considered an ensemble composed of graphs synthesized according to a recently-proposed mechanism \cite{communitystructure_bmc2014}, designed to construct a network with specified modularity and degree profiles. Notably, we reproduced the modularity and degree values from the actual protein contact networks herein considered.
Results demonstrate that modularity, when hardcoded into the networks, does not explain the actual architecture of proteins.
In fact, we concluded that modularity should be considered as an emergent property of such networks, which is suitably optimized in a trade-off with the conflicting feature of path efficiency.
In our model, modularity emerged from the peculiar PCNs mesoscopic wiring obtained from their empirical contact distribution at increasing distance length: a simple linear decrease in contact frequency at increasing sequence distance does not allow to reach the typical modularity of real proteins.
The fine tuning of long-range contacts allows for directly intervening on both modularity and path efficiency balance, so confirming the crucial importance of long-range contacts in folding process \cite{vendruscolo2001three,chiti1999mutational}.

A sound generative mechanism for protein contact networks is of utmost importance in current researches in protein science.
In future works, in addition to the improvement of the herein proposed method, we also plan to tackle the problem in a data-driven generative learning scenario, for instance using generative (deep) neural networks \cite{krizhevsky2012imagenet}.
The possibility to learn in a data-driven fashion an effective and sound model for protein contact networks would allow to easily generalize other instances of such networks. This perspective could be interesting also for protein engineering purposes \cite{currin2015synthetic}.
The theoretical study of networks promises to pave the way for the discovery of universal principles at the basis of biological organization as well as instructing the generation of technological devices.

\appendix
\section{Methods}
\label{sec:methods}

\subsection{The proposed generative method for synthesizing protein contact networks}
\label{sec:our_method}

In this section we describe the proposed generative method for PCN. Instead of merely presenting the mechanism itself, we first discuss an important fact related to the distance on sequence of amino acid residues and their relations on the native 3D structure of proteins.
This initial discussion, in our opinion, is relevant for the purpose of designing and, most importantly justifying, a generative mechanism for PCN.

\begin{figure}[ht!]
\centering

\subfigure[SR contacts.]{
\includegraphics[viewport=0 0 340 243,scale=0.62,keepaspectratio=true]{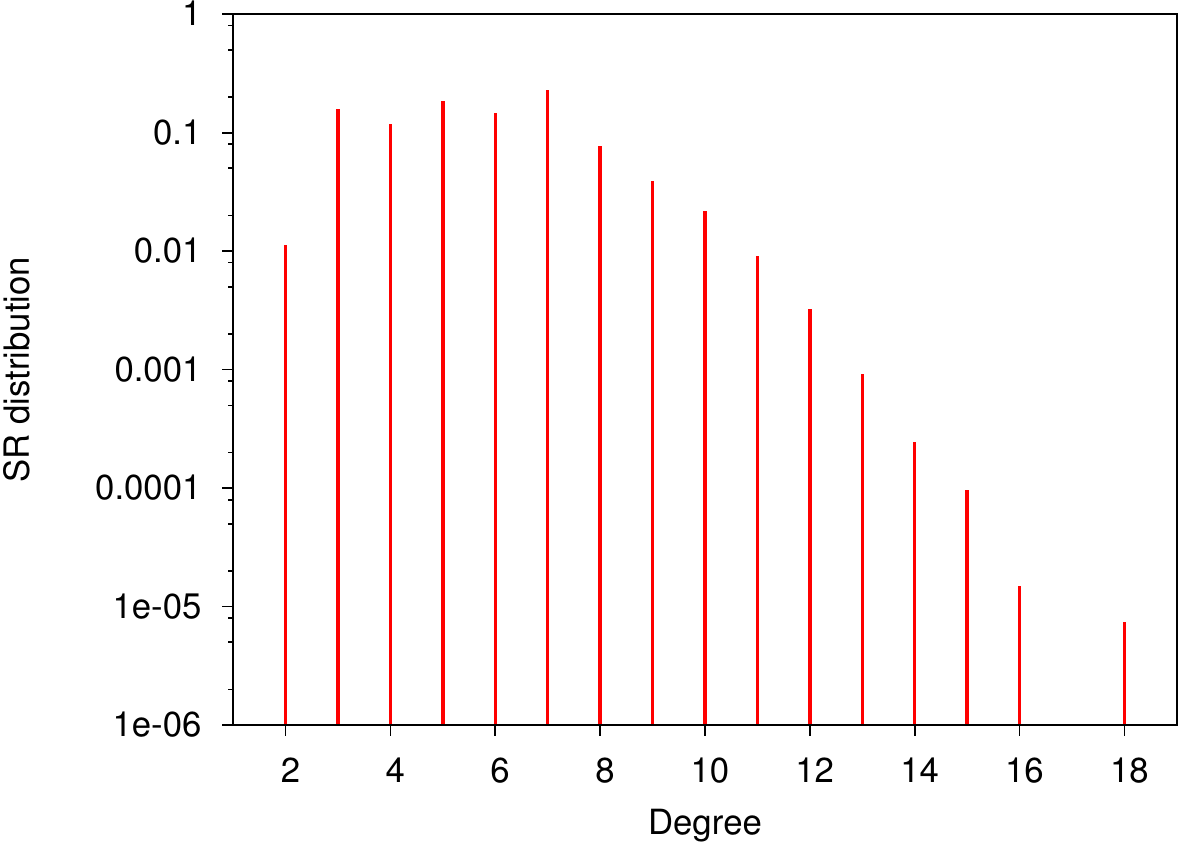}
\label{fig:sr_distr}}
~
\subfigure[Distribution of LR contacts.]{
\includegraphics[viewport=0 0 339 243,scale=0.62,keepaspectratio=true]{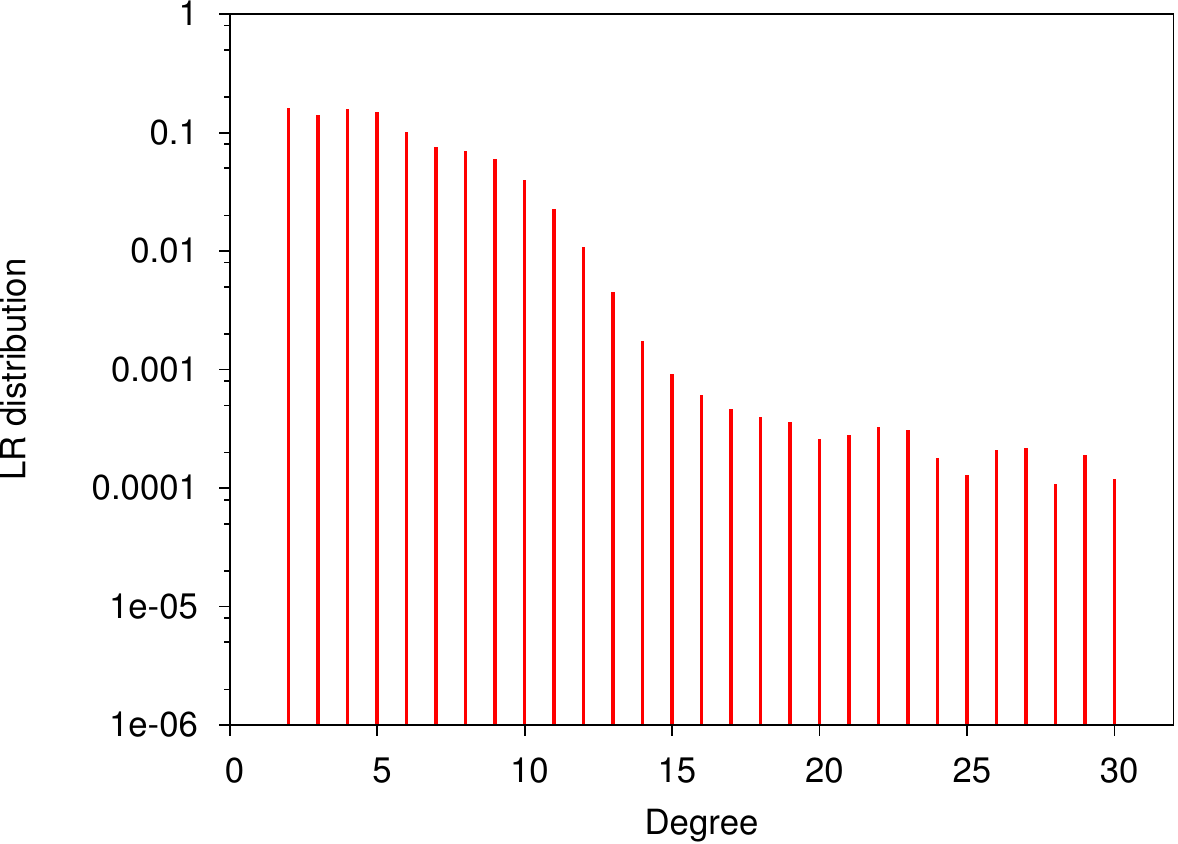}
\label{fig:lr_distr}}

\caption{Degree distribution of SR (\ref{fig:sr_distr}) and LR (\ref{fig:lr_distr}) contacts. Both distributions are provided in lin-log plots to improve visualization. SR contacts are determined by considering a distance on the sequence lower than or equal to 12 residues.}
\label{fig:sr_lr_contacts}
\end{figure}

Native contacts in folded proteins are in one way or another constrained by the covalent bonds due to the backbone.
Therefore, a first interesting question that one would ask when designing a generative mechanism is ``what is the effect of the backbone on the degree distribution of a PCN''. To provide an answer to such a question, we first define the notion of short range (SR) and long range (LR) contacts, that is, native contacts whose residues are, respectively, close and distant on the sequence (backbone). We chose 12 residues as threshold for SR contacts \cite{paci2012structural}.
Fig. \ref{fig:sr_lr_contacts} shows the two separate degree distributions elaborated from the considered ensemble of varying-size PCNs.
SR contacts denote a clearly different distribution with respect to those that are LR; the latter is vaguely compatible with a power-law.

\begin{figure}[ht!]
\centering

\subfigure[JW0058 modules/links organization.]{
\includegraphics[viewport=0 0 1321 701,scale=0.34,keepaspectratio=true]{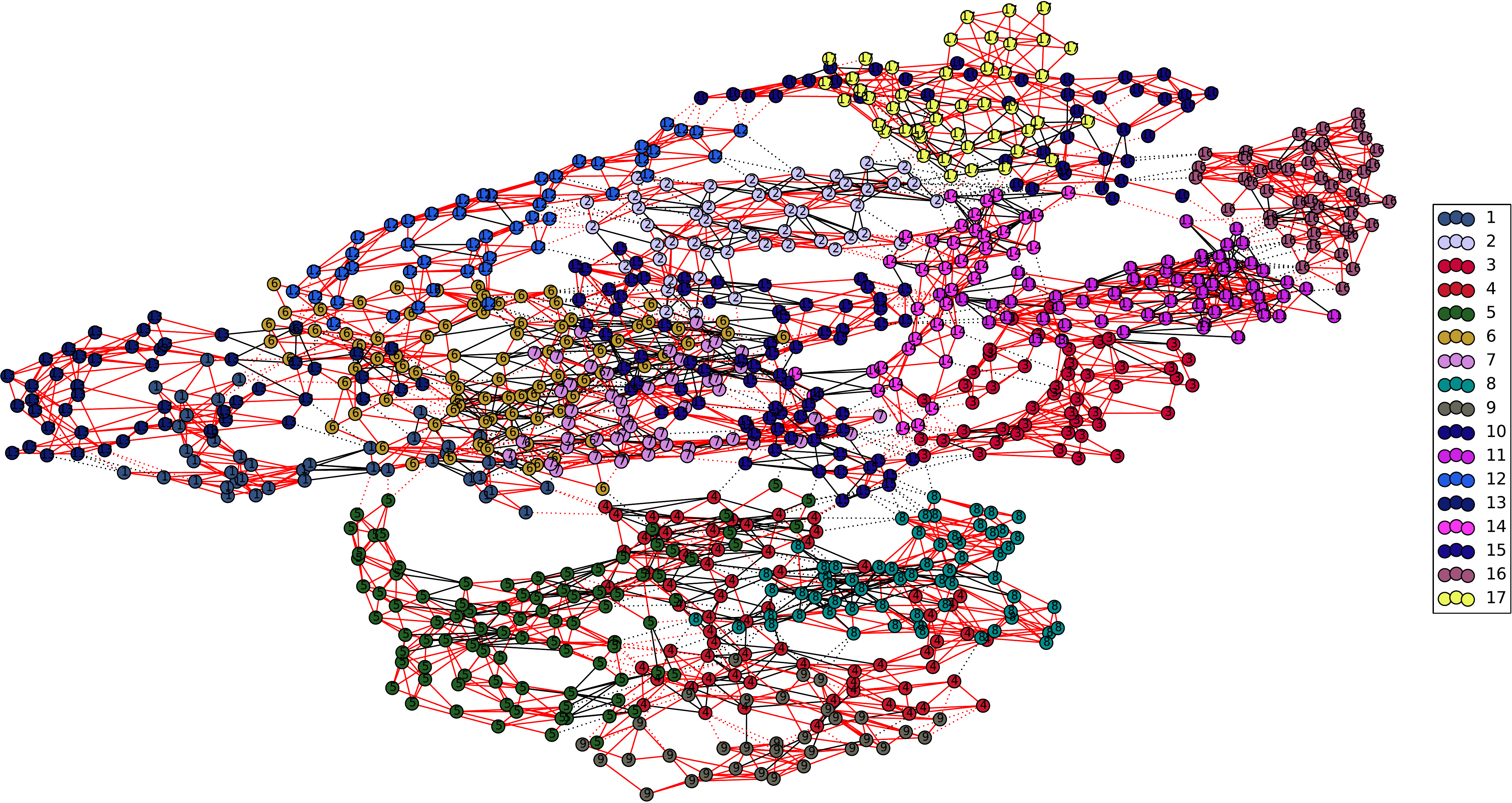}
\label{fig:JW0058__modules}}
~
\subfigure[JW0179 modules/links organization.]{
\includegraphics[viewport=0 0 1328 675,scale=0.34,keepaspectratio=true]{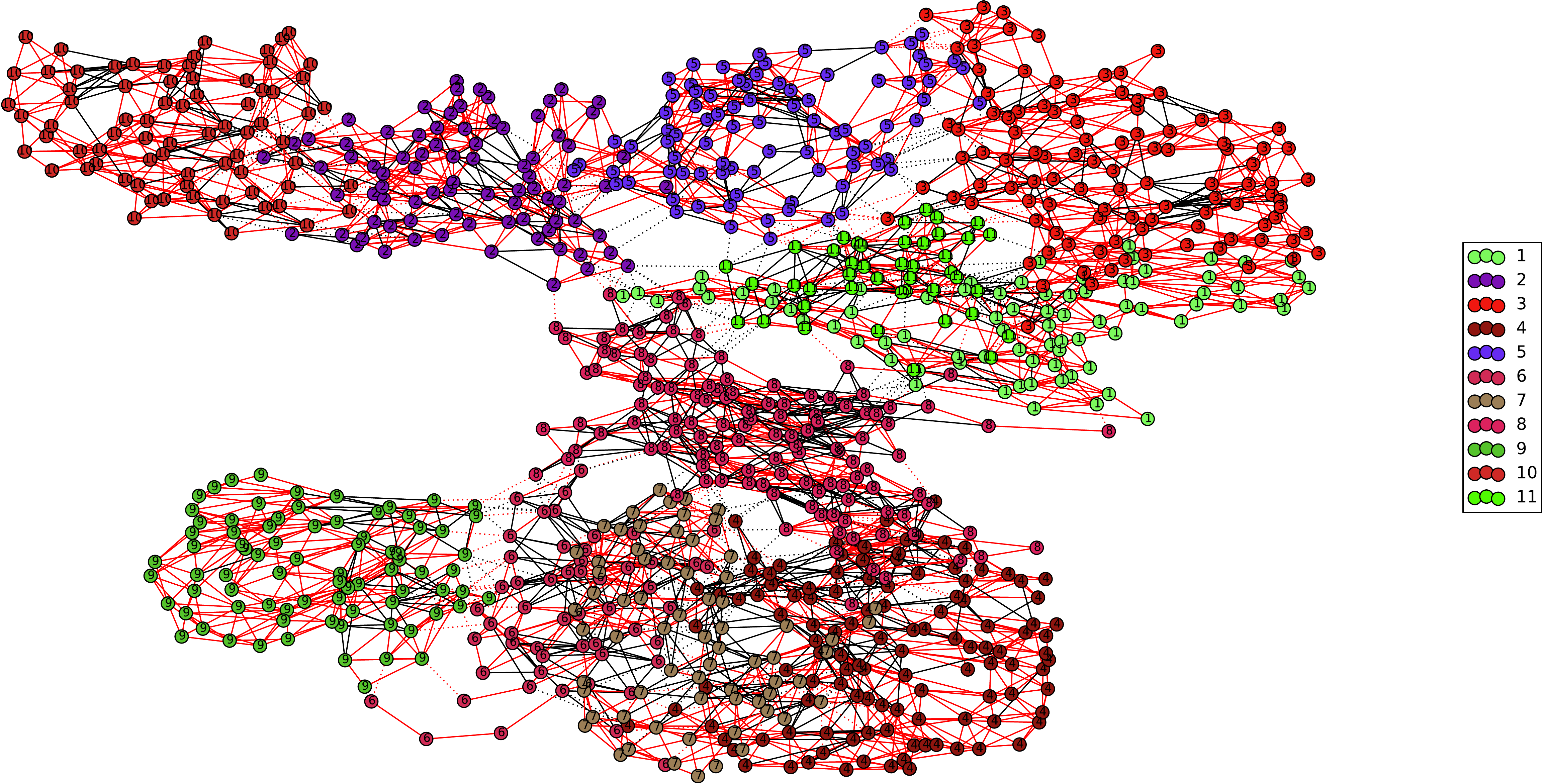}
\label{fig:JW0179__modules}}

\caption{Classification of contacts by considering the SR/LR typology and the intra/inter module arrangement. The partition is derived by using the maximum modularity criterion. Vertex assignment to modules is represented using different colors; numerical module identifies are drawn in the legend and in the corresponding vertex labels. Solid lines denote intra-module links while dashed lines inter-module links. Black links denote LR contacts, while red links are SR. Please note that the length of the links in the figures respects the actual Euclidean distances of contacts. The assumption that LR contacts are mostly inter-module links (and accordingly, SR contacts are mostly intra-module links) seems to be disproved by those examples.}
\label{fig:module_links}
\end{figure}

Considering this fact and that PCNs do posses a modular architecture, one would be tempted to postulate a striking rule such as ``SR contacts are intra-module while LR are inter-module links''. If this rule would be correct, it would be possible to design a generative mechanism accordingly, e.g., by connecting intra-module and inter-module links according to their specific (empirical) distributions; although the number of modules should be defined a priori.
Nevertheless, such a possibility seems to be weakly supported by the following test.
In Fig. \ref{fig:module_links} we show a graphical representations of two PCNs, denoted as ``JW0058'' and ``JW0179''.
Those two networks contain roughly the same number of amino acid residues (around a thousand); JW0058 is made of two chains while JW0179 is derived from a single-chain polymer.
To verify the above stated hypothesis, we need to consider a suitable criterion to generate a partition (i.e., to group the vertices into modules).
In our case, we have chosen to use the partition having maximum modularity as computed with the algorithm presented in Ref. \cite{blondel2008fast}.
Results in Fig. \ref{fig:module_links} demonstrate that intra-module links (solid lines) are SR (drawn in red), in both cases, only around 55\% of the times.
This fact -- that has been verified for a larger number of PCN -- suggests to reconsider the possibility to follow such a SR/LR contacts characterization with respect to intra/inter module links.
In addition, we found in our data that there is no trivial relation among the distance on sequence and the Euclidean distance among residues in the 3D space ($r\simeq0.162$, details not shown here).
These facts find confirmation by considering the enormous research effort in predicting native contacts starting from the sequence \cite{morcos2011direct,de2013emerging,marks2011protein,hopf2012three,baldassi2014fast,ekeberg2013improved,kamisetty2013assessing,skwark2014improved,morcos2013coevolutionary,C3CP55275F}.

\begin{figure}[ht!]
\centering
 \includegraphics[viewport=0 0 351 243,scale=0.7,keepaspectratio=true]{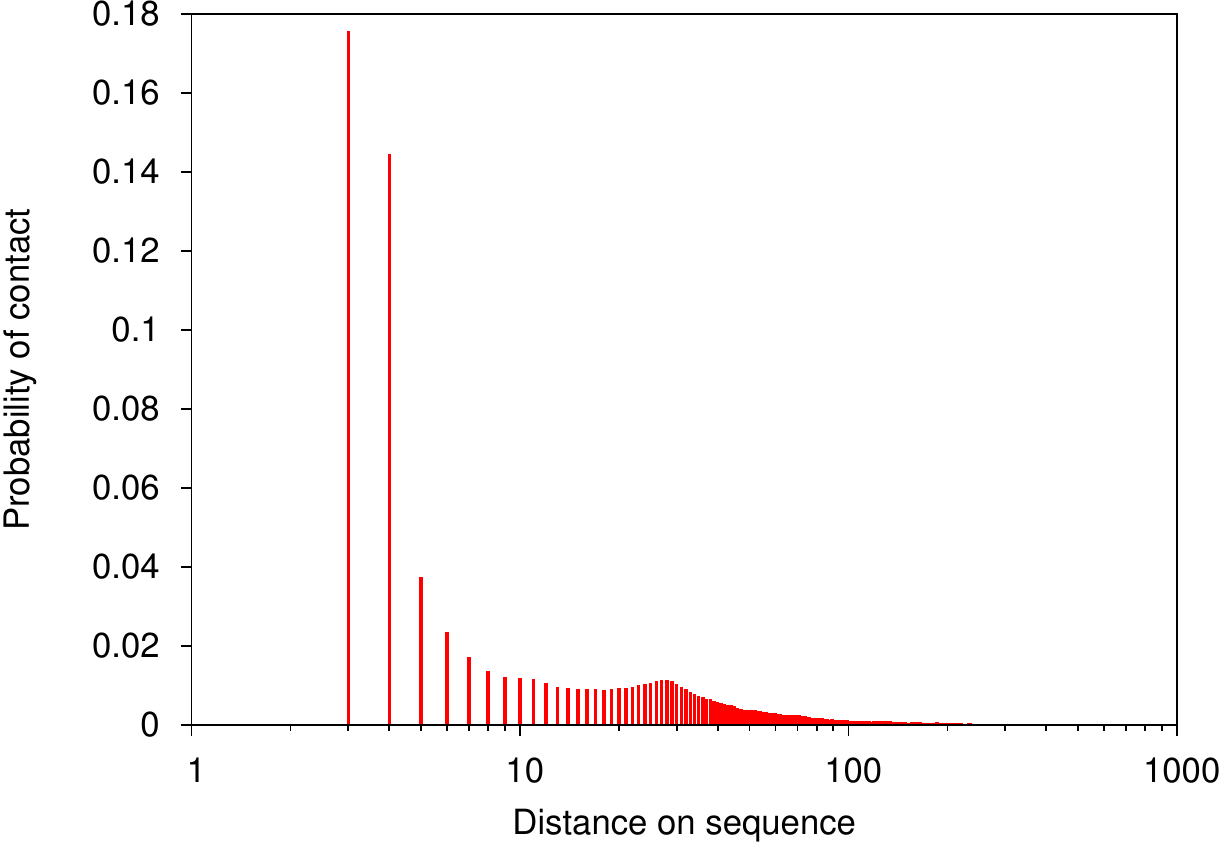}
 \caption{Empirical distribution of contacts considering the distance on sequence (backbone distance). The distribution is elaborated from the entire ensemble of PCN without considering contacts at distance one and two on the sequence; backbone contacts are added deterministically.}
\label{fig:empirical_distribution}
\end{figure}

Let us describe the proposed generative mechanism.
Algorithm \ref{alg:our_method} conveys the pseudo-code of the procedure. The algorithm takes inspiration from the mechanism introduced by \citet{bartoli2007effect}.
Firstly, edges are deterministically added among any two residues at distance two on the sequence. This provides the definition for backbone contacts.
The main difference with \citet{bartoli2007effect} is that, to add all remaining non-backbone contacts, we use the empirical distribution shown in Fig. \ref{fig:empirical_distribution} instead of a linear function of the sequence distance.
As shown in the results, this straightforward modification resulted in a considerable improvement under many aspects.
\begin{algorithm}[h!]\footnotesize
\caption{Pseudo-code of the proposed generative algorithm.}
\label{alg:our_method}
\begin{algorithmic}[1]
\REQUIRE Number of vertices, $n$, and edges, $m$
\ENSURE A graph $G=(\mathcal{V}, \mathcal{E})$ with $n=|\mathcal{V}|$ and $m=|\mathcal{E}|$
\STATE Add $n$ vertices in $\mathcal{V}$ with unique, progressive, numerical identifiers
\STATE Add backbone contacts in $\mathcal{E}$: connect all vertices $v_i$ and $v_j$ for which $|i-j|=2$
\STATE Loop to add all remaining non-backbone contacts $\curvearrowright$
\WHILE{$|\mathcal{E}|<m$}
\STATE Select two non-connected vertices $v_i$ and $v_j$ with probability $p(|i-j|)$ given by their distance $|i-j|$ according to the empirical distribution in Fig. \ref{fig:empirical_distribution}
\STATE Add the undirected edge $e=(v_i, v_j)$ in $\mathcal{E}$
\ENDWHILE
\RETURN $G=(\mathcal{V}, \mathcal{E})$
\end{algorithmic}
\end{algorithm}

We now introduce the second step of the proposed generative mechanism, which implements the edge reconfiguration.
Given a graph $G=(\mathcal{V}, \mathcal{E})$, the herein introduced reconfiguration step is primarily meant to lower the small-world signature in $G$.
This is done by iteratively rewiring edges in $\mathcal{E}$ according to their edge-betweenness value.
The pseudo-code of the edge reconfiguration algorithm in delivered by Algorithm \ref{alg:edge_rec}.
The reconfigured graph $\hat{G}$ is obtained at the end of the reconfiguration loop. Please note that we insure connectedness for all $\hat{G}$.
The loss of small-world signature in $\hat{G}$ is primarily verified with the ASP increase (see Fig. \ref{fig:jw0058_lmgr__ASP} for an example), which is a consequence of the targeted rewiring of edges with maximum edge-betweenness.
\begin{algorithm}[h!]\footnotesize
\caption{Pseudo-code of the proposed edge reconfiguration algorithm.}
\label{alg:edge_rec}
\begin{algorithmic}[1]
\REQUIRE A graph $G=(\mathcal{V}, \mathcal{E})$ with $n=|\mathcal{V}|$ and $m=|\mathcal{E}|$
\ENSURE A modified graph $\hat{G}=(\mathcal{V}, \mathcal{E})$ with $n=|\mathcal{V}|$ and $m=|\mathcal{E}|$
\LOOP
\STATE Calculate the edge-betweenness measure for all edges in $\mathcal{E}$
\STATE Let $e_{\mathrm{max}}$ be the edge with maximum edge-betweenness. Remove $e_{\mathrm{max}}$ from $\mathcal{E}$
\STATE Select two non-connected vertices $v_i$ and $v_j$ with probability $p(|i-j|)$ given by their distance $|i-j|$ according to the empirical distribution in Fig. \ref{fig:empirical_distribution}
\STATE Add the undirected edge $e=(v_i, v_j)$ in $\mathcal{E}$
\ENDLOOP{ when stop criterion is met}
\RETURN $\hat{G}=(\mathcal{V}, \mathcal{E})$
\end{algorithmic}
\end{algorithm}

\clearpage
\subsection{Graph characterization and heat kernel}
\label{sec:graph_characterization}

In this appendix we provide essential technical details regarding the graph characterizations used in this paper to study the four ensembles of variable-size graphs representing proteins.

The first characterization employs classical topological descriptors, which include statistics of the degrees/shortest paths and also some elaborations of the graph spectra.
We consider the modularity (MOD) \cite{newman2006modularity,blondel2008fast} for quantifying the presence of a global community structure -- please note that we consider the value associated to the partition with maximum modularity. Then we consider the average closeness centrality (ACC), average shortest path (ASP), and the average clustering coefficient (ACL) \cite{costa2007characterization}; the energy (EN) and Laplacian energy (LEN) of the corresponding spectra  \cite{gutman2006laplacian}; the ambiguity (A) \cite{Livi_ga_2013}, which expresses the degree of irregularity of the topology; and finally the 2-order R\'{e}nyi entropy of a stationary Markovian random walk (H) \cite{Dehmer201157}.

In what follows, we describe the heat kernel and the derived invariants used to characterize the considered ensembles.
The following material is principally reorganized from Ref. \cite{mixbionets1__arxiv}.
Let $G=(\mathcal{V}, \mathcal{E})$ be a graph with $n=|\mathcal{V}|$ vertices and $m=|\mathcal{E}|$ edges.
Let $\mathbf{A}^{n\times n}$ be the adjacency matrix defined as $A_{ij}=1$ if there is an edge between vertices $v_i,v_j\in\mathcal{V}$; $A_{ij}=0$ otherwise. Let us define the degree of a vertex $v_i$ as $\mathrm{deg}(v_i)=\sum_{j=1}^{n} A_{ij}$. In addition, let us define $\mathbf{D}$ as a diagonal matrix of degree: $D_{ii}=\mathrm{deg}(v_i)$.
Let $\mathbf{L}$ be the Laplacian matrix given by $\mathbf{L}=\mathbf{D}-\mathbf{A}$; the normalized Laplacian matrix as $\hat{\mathbf{L}}=\mathbf{D}^{-1/2}\mathbf{L}\mathbf{D}^{-1/2}$.
As a consequence, $\hat{\mathbf{L}}$ is symmetric and positive semi-definite and therefore it has non-negative eigenvalues.
The eigendecomposition of the Laplacian is expressed as $\hat{\mathbf{L}}=\Phi \Lambda \Phi^{T}$, where $\Lambda$ is the diagonal matrix containing the eigenvalues arranged as $0=\lambda_1\leq \lambda_2\leq ...\leq \lambda_n\leq 2$; $\Phi$ contains the corresponding (unitary) eigenvectors as columns.

The heat equation \cite{Xiao:2009:GCH:1563046.1563099} associated to $\hat{\mathbf{L}}$ is given by
\begin{equation}
\label{eq:heat_equation}
\frac{\partial \mathbf{H}_{t}}{\partial t} = -\hat{\mathbf{L}}\mathbf{H}_{t},
\end{equation}
where $\mathbf{H}_{t}$ is a doubly-stochastic $n\times n$ matrix, called heat matrix, and $t$ is the time variable.
Eq. \ref{eq:heat_equation} describes the diffusion of heat/information across the graph over time. Being doubly-stochastic, the heat matrix possesses a uniform stationary distribution.
It is well-known that the solution to (\ref{eq:heat_equation}) is
\begin{equation}
\mathbf{H}_{t} = \exp(-\hat{\mathbf{L}}t),
\end{equation}
which can be solved by exponentiating the spectrum of $\hat{\mathbf{L}}$:
\begin{equation}
\mathbf{H}_{t} = \Phi \exp(-\Lambda t) \Phi^{T} = \sum_{i=1}^{n} \exp(-\lambda_i t)\phi_i\phi_i^{T}.
\end{equation}

The heat trace (HT) of $\mathbf{H}_{t}$ is an invariant feature that is given by
\begin{equation}
\label{eq:heat_trace}
\mathrm{HT}(t) = \mathrm{Tr}(\mathbf{H}_{t}) = \sum_{i=1}^{n} \exp(-\lambda_i t),
\end{equation}
which thus takes into account only the eigenvalues of $\hat{\mathbf{L}}$.
The heat content (HC) of $\mathbf{H}_t$ is defined by considering also the eigenvectors of $\hat{\mathbf{L}}$:
\begin{align}
\label{eq:heat_content}
\mathrm{HC}(t) = \sum_{u\in\mathcal{V}} \sum_{u\in\mathcal{V}} \mathbf{H}_{t}(u, v) = \sum_{u\in\mathcal{V}} \sum_{u\in\mathcal{V}} \sum_{i=1}^{n} \exp(-\lambda_i t)\phi_i(v)\phi_i(u),
\end{align}
where with $\phi_i(v)$ we indicate the value related to the vertex $v$ in the \textit{i}th eigenvector.

Eq. \ref{eq:heat_content} can be described in terms of power series expansion,
\begin{equation}
\label{eq:heat_coeff}
\mathrm{HC}(t) = \sum_{m=0}^{\infty} q_m t^m.
\end{equation}

By using the McLaurin series for the exponential function, we have
\begin{equation}
\exp(-\lambda_i t)=\sum_{m=0}^{\infty} \frac{(-\lambda_i)^{m} t^m}{m!},
\end{equation}
which substituted in Eq. \ref{eq:heat_content} gives:
\begin{align}
\label{eq:heat_content2}
\mathrm{HC}(t) = \sum_{u\in\mathcal{V}} \sum_{u\in\mathcal{V}} \sum_{i=1}^{n} \exp(-\lambda_i t)\phi_i(v)\phi_i(u) = \sum_{m=0}^{\infty}\sum_{u\in\mathcal{V}} \sum_{u\in\mathcal{V}} \sum_{i=1}^{n} \phi_i(v)\phi_i(u) \frac{(-\lambda_i)^{m} t^m}{m!}.
\end{align}

The $q_m$ coefficients in (\ref{eq:heat_coeff}) are graph invariants (called heat content invariants, HCI) that can be calculated in closed-form by using Eqs. \ref{eq:heat_coeff} and \ref{eq:heat_content2}:
\begin{equation}
\label{eq:hci}
q_m = \displaystyle\sum_{i=1}^{n} \left( \sum_{u\in\mathcal{V}} \phi_{i}(u) \right)^2 \frac{(-\lambda_i)^m}{m!}.
\end{equation}

In what follows, we provide an argument to characterize the heat trace (\ref{eq:heat_trace}) as a property of a homogeneous ensemble of graphs.
The heat trace (\ref{eq:heat_trace}) of a graph $G=(\mathcal{V}, \mathcal{E})$, with $n = |\mathcal{V}|$, can be expressed as
\begin{equation}
\label{eq:httv}
\text{HT}_G(t) = \sum_{i=1}^n \exp(- \lambda_i t ) = 1 + \sum_{i=2}^n \exp(- \lambda_i t ) .
\end{equation}
where $\lambda_i$ are the eigenvalues of the normalized Laplacian of $G$.
Let us define an ensemble $\mathcal{C}$ of graphs, in which all graphs share a common characteristic spectral density. Such spectra can be synthetically described by considering the spectral density of $\mathcal{C}$. Accordingly, we can consider the eigenvalues as i.i.d. random variables, $\tilde{\lambda}_i$, assuming values according to the spectral density of the ensemble; note that $\tilde{\lambda}_1$ assumes deterministically the value 0.
The HT of a generic graph $G \in \mathcal{C}, n=|\mathcal{V}|,$ can be written as:
\begin{equation}
\label{eq:ht_graph}
\text{HT}_G (t; n) = 1 + \sum_{i=2}^n \exp(- \tilde{\lambda}_i t ) = 1 + \sum_{i=2}^n \exp(- \tilde{\lambda} t).
\end{equation}

The last step in Eq. \ref{eq:ht_graph} is carried out by considering that, since the $\tilde{\lambda}_i$ are assumed as i.i.d., their values can be expressed as $n$ realizations of a single random variable, $\tilde{\lambda}$, characterized by the same probability density function.
For a fixed value of time $t$, we can define the ensemble HT, $\text{HT}_\mathcal{C}(n; t)$, as the mean HT over all graphs of the ensemble $\mathcal{C}$ with varying size $n$. This quantity is given by:
\begin{equation}
\text{HT}_\mathcal{C}(n; t) = \langle \text{HT}_G(t; n) \rangle_\mathcal{C} = 1 + \sum_{i=2}^n \langle \exp(- \tilde{\lambda} t ) \rangle_\mathcal{C} = 1 + (n-1) \langle \exp(- \tilde{\lambda} t ) \rangle_\mathcal{C}.
\end{equation}

Hence, $\text{HT}_\mathcal{C}(n; t)$ can be expressed as a linear function of the graph size
\begin{equation}
\label{eq:ensemble_ht}
\text{HT}_\mathcal{C}(n; t) = 1 - \alpha_\mathcal{C}(t) + \alpha_\mathcal{C}(t) \cdot n \simeq \alpha_\mathcal{C}(t) \cdot n,
\end{equation}
where $\alpha_\mathcal{C}(t) = \langle \exp{(- \tilde{\lambda} t)} \rangle_\mathcal{C}\in[0, 1]$ is a time-dependent slope that is characteristic for the entire ensemble $\mathcal{C}$.

\bibliographystyle{abbrvnat}
\bibliography{/home/lorenzo/University/Research/Publications/Bibliography.bib}
\end{document}